\documentclass[]{article}
\makeatletter\if@twocolumn\PassOptionsToPackage{switch}{lineno}\else\fi\makeatother

\usepackage{amsmath,amsfonts,amsbsy,amssymb,tabulary,graphicx,times,caption,fancyhdr,amsthm, rotating,enumerate,bm}
\usepackage[utf8]{inputenc}
\usepackage[paperheight=10in,paperwidth=6.5in,margin=2cm,headsep=.5cm,top=2cm,headheight=1cm]{geometry}
\renewenvironment{abstract} {\vspace*{-1pc}\trivlist\item[]\leftskip\oupIndent\hrulefill\par\vskip4pt\noindent\textbf{\abstractname}\mbox{\null}\\\relax}{\par\noindent\hrulefill\endtrivlist} 
 \date{}
\usepackage{caption,booktabs}

\makeatletter\def\oupIndent{1pt}
\def\author#1{\gdef\@author{\hskip-\dimexpr(\tabcolsep)\hskip\oupIndent\parbox{\dimexpr\textwidth-\oupIndent}{\centering\bfseries#1}}}
\def\title#1{\gdef\@title{\centering\bfseries\ifx\@articleType\@empty\else\@articleType\\\fi#1}}
\let\@articleType\@empty \def\articletype#1{\gdef\@articleType{{\normalfont\itshape#1}}}
\fancypagestyle{headings}{\fancyhf{}\fancyhead[C]{\RunningHead}\fancyhead[R]{\thepage}}
\makeatother

\usepackage{url,multirow,morefloats,floatflt,cancel,tfrupee}
\makeatletter

\AtBeginDocument{\@ifpackageloaded{textcomp}{}{\usepackage{textcomp}}}
\makeatother
\usepackage{colortbl}
\usepackage{xcolor}
\usepackage{pifont}
\usepackage[nointegrals]{wasysym}
\makeatletter

\def\mcWidth#1{\csname TY@F#1\endcsname+\tabcolsep}

\def\cAlignHack{\rightskip\@flushglue\leftskip\@flushglue\parindent\z@\parfillskip\z@skip}
\def\rAlignHack{\rightskip\z@skip\leftskip\@flushglue \parindent\z@\parfillskip\z@skip}

\@ifundefined{etal}{}{}

\usepackage{ifxetex}
\ifxetex\else\if@twocolumn\@ifpackageloaded{stfloats}{}{\usepackage{dblfloatfix}}\fi\fi

\AtBeginDocument{
\expandafter\ifx\csname eqalign\endcsname\relax
\def\eqalign#1{\null\vcenter{\def\\{\cr}\openup\jot\m@th
  \ialign{\strut$\displaystyle{##}$\hfil&$\displaystyle{{}##}$\hfil
      \crcr#1\crcr}}\,}
\fi
}

\AtBeginDocument{%
  \@ifpackageloaded{endfloat}%
   {\renewcommand\efloat@iwrite[1]{\immediate\expandafter\protected@write\csname efloat@post#1\endcsname{}}}{\newif\ifefloat@tables}%
}%

\def\BreakURLText#1{\@tfor\brk@tempa:=#1\do{\brk@tempa\hskip0pt}}
\let\lt=<
\let\gt=>
\def\processVert{\ifmmode|\else\textbar\fi}

\@ifundefined{subparagraph}{
\def\subparagraph{\@startsection{paragraph}{5}{2\parindent}{0ex plus 0.1ex minus 0.1ex}%
{0ex}{\normalfont\small\itshape}}%
}{}

\newcommand\role[1]{\unskip}
\newcommand\aucollab[1]{\unskip}

\@ifundefined{tsGraphicsScaleX}{\gdef\tsGraphicsScaleX{1}}{}
\@ifundefined{tsGraphicsScaleY}{\gdef\tsGraphicsScaleY{.9}}{}
\def\checkGraphicsWidth{\ifdim\Gin@nat@width>\linewidth
	\tsGraphicsScaleX\linewidth\else\Gin@nat@width\fi}

\def\checkGraphicsHeight{\ifdim\Gin@nat@height>.9\textheight
	\tsGraphicsScaleY\textheight\else\Gin@nat@height\fi}

\def\fixFloatSize#1{}
\let\ts@includegraphics\includegraphics

\def\inlinegraphic[#1]#2{{\edef\@tempa{#1}\edef\baseline@shift{\ifx\@tempa\@empty0\else#1\fi}\edef\tempZ{\the\numexpr(\numexpr(\baseline@shift*\f@size/100))}\protect\raisebox{\tempZ pt}{\ts@includegraphics{#2}}}}

\AtBeginDocument{\def\includegraphics{\@ifnextchar[{\ts@includegraphics}{\ts@includegraphics[width=\checkGraphicsWidth,height=\checkGraphicsHeight,keepaspectratio]}}}

\DeclareMathAlphabet{\mathpzc}{OT1}{pzc}{m}{it}

\def\URL#1#2{\@ifundefined{href}{#2}{\href{#1}{#2}}}

\def\UrlOrds{\do\*\do\-\do\~\do\'\do\"\do\-}%
\g@addto@macro{\UrlBreaks}{\UrlOrds}

\edef\fntEncoding{\f@encoding}

\makeatother

\newif\ifmultipleabstract\multipleabstractfalse%

\makeatletter
\newcommand*{\centerfloat}{%
  \parindent \z@
  \leftskip \z@ \@plus 1fil \@minus \textwidth
  \rightskip\leftskip
  \parfillskip \z@skip}
\makeatother

\AtBeginDocument{\let\citet\cite\let\citep\cite}
\usepackage[numbers,sort&compress,numbers,sort&compress]{natbib}

\usepackage[T1]{fontenc}

\makeatletter
\renewcommand{\boxed}[1]{\text{\fboxsep=.2em\fbox{\m@th$\displaystyle#1$}}}
\makeatother

\makeatletter
\newenvironment{breakablealgorithm}
  {
   \begin{center}
     \refstepcounter{algorithm}
     \hrule height.8pt depth0pt \kern2pt
     \renewcommand{\caption}[2][\relax]{
       {\raggedright\textbf{\fname@algorithm~\thealgorithm} ##2\par}%
       \ifx\relax##1\relax 
         \addcontentsline{loa}{algorithm}{\protect\numberline{\thealgorithm}##2}%
       \else 
         \addcontentsline{loa}{algorithm}{\protect\numberline{\thealgorithm}##1}%
       \fi
       \kern2pt\hrule\kern2pt
     }
  }{
     \kern2pt\hrule\relax
   \end{center}
  }
\makeatother

\usepackage[flushleft]{threeparttablex}
\usepackage{setspace}
\usepackage{algorithm,algpseudocode}
\usepackage{algpseudocode}
\usepackage{yhmath}
\usepackage{float}
\usepackage{enumitem}
\setlist[enumerate]{itemsep=0mm,leftmargin=5mm}
\begin{document}

\nocite{*}

\title{Sensitivity analysis methods for outcome missingness using substantive-model-compatible multiple imputation and their application in causal inference}

\author{\textbf{\fontsize{10pt}{16.4pt}\selectfont{Jiaxin Zhang\textsuperscript{1,2,*}, S. Ghazaleh Dashti\textsuperscript{2,1}, John B. Carlin\textsuperscript{2,1}, Katherine J. Lee\textsuperscript{2,1}, Jonathan W. Bartlett\textsuperscript{3} and Margarita Moreno-Betancur\textsuperscript{1,2}}}~\\
{\small\normalfont \textsuperscript{1} Clinical Epidemiology and Biosatistics Unit, Department of Paediatrics, University of Melbourne, Australia\\
\textsuperscript{2} Clinical Epidemiology and Biosatistics Unit, Murdoch Children’s Research Institute, Australia\\
\textsuperscript{3} Department of Medical Statistics, London School of Hygiene \& Tropical Medicine, UK\\
\textsuperscript{*}  jiaxizhang1@student.unimelb.edu.au~}}
\def\RunningHead{{}}

\maketitle 
\def\keywordstitle{Wordcount}

\abstract{
When using multiple imputation (MI) for missing data, maintaining compatibility between the imputation model and substantive analysis is important for avoiding bias. For example, some causal inference methods incorporate an outcome model with exposure-confounder interactions that must be reflected in the imputation model. Two approaches for compatible imputation with multivariable missingness have been proposed: Substantive-Model-Compatible Fully Conditional Specification (SMCFCS) and a stacked-imputation-based approach (SMC-stack). If the imputation model is correctly specified, both approaches are guaranteed to be unbiased under the ``missing at random’’ assumption. However, this assumption is violated when the outcome causes its own missingness, which is common in practice. In such settings, sensitivity analyses are needed to assess the impact of alternative assumptions on results. An appealing solution for sensitivity analysis is delta-adjustment using MI, specifically ``not-at-random’’ (NAR)FCS. However, the issue of imputation model compatibility has not been considered in sensitivity analysis, with a naïve implementation of NARFCS being susceptible to bias. To address this gap, we propose two approaches for compatible sensitivity analysis when the outcome causes its own missingness. The proposed approaches, NAR-SMCFCS and NAR-SMC-stack, extend SMCFCS and SMC-stack, respectively, with delta-adjustment for the outcome. We evaluate these approaches using a simulation study that is motivated by a case study, to which the methods were also applied. The simulation results confirmed that a naïve implementation of NARFCS produced bias in effect estimates, while NAR-SMCFCS and NAR-SMC-stack were approximately unbiased. The proposed compatible approaches provide promising avenues for conducting sensitivity analysis to missingness assumptions in causal inference.
}

\def\keywordstitle{Keywords}

\def\keywordstitle{Keymessages}

\clearpage
\section{Introduction}\label{sec: Introduction}
Missing data is a common issue in epidemiological studies and can make estimating the parameters of interest, such as the average causal effect (ACE), challenging \cite{Causal2020,lee2021framework,zhang2024recoverability}. Multiple imputation (MI) is a widely used tool for handling missing data in which missing values are imputed multiple times from a specified imputation model. The substantive analysis is conducted on each imputed dataset separately; then, the results are pooled into a final estimate \cite{rubin2004multiple,van2007multiple,carpenter2023multiple}. It is widely recognised that the imputation model must accommodate, or be compatible with, the model(s) used in the substantive analysis, or else bias can arise if the imputation model makes parametric assumptions that conflict with those of the analysis, e.g. omission of an interaction term \cite{meng1994multiple,liu2014stationary}. 

Recently, two MI approaches have been developed that facilitate the accommodation of the substantive analysis when this consists of fitting a regression model, and there is missingness in more than one variable required for analysis (referred to as multivariable missingness). The Substantive Model Compatible Fully Conditional Specification (SMCFCS) approach \cite{bartlett2015multiple} is a compatible version of the widely used FCS imputation paradigm \cite{rubin2004multiple,van2007multiple,carpenter2023multiple}. In SMCFCS, each univariate imputation model is factorised as the product of the distribution of non-outcome variables and the substantive outcome model \cite{bartlett2015multiple}, and is imputed sequentially within an FCS procedure. Alternatively, the stacked-imputation-based approach, referred to as SMC-stack, imputes non-outcome variables independently from the outcome multiple times, following which a new dataset is obtained by stacking the imputed datasets. The substantive regression model is then fitted once to this stacked dataset using weights that are proportional to the substantive model distribution \cite{beesley2021stacked}.

Under the ``missing at random’’ (MAR) assumption \cite{seaman2013meant}, both approaches have been shown to perform well in terms of reducing the bias that would arise when using imputation models that ignore compatibility and as a result are misspecified. We refer to these approaches as MAR-SMC approaches. However, the MAR missingness assumption is opaque and of questionable relevance in multivariable missingness settings \cite{moreno2018canonical,lee2023assumptions,zhang2024recoverability}. Importantly, the violation of the MAR assumption is common in practice, where incomplete variables often cause their own missingness. In the context of causal inference, previous studies have shown that the ACE is not non-parametrically identifiable, or ``recoverable’’, when the outcome causes its own missingness, which is the scenario of focus in this manuscript \cite{mohan2013graphical,zhang2024recoverability}. Further, simulation studies have shown that standard MI (commonly characterised as ``MI under an MAR assumption’’) returns estimates with high bias under such missingness mechanisms \cite{moreno2018canonical,zhang2024recoverability}. Therefore, as suggested in previous studies \cite{moreno2018canonical,zhang2024recoverability}, sensitivity analysis should be conducted to assess the robustness of results under missingness mechanisms where it is assumed that the outcome might have caused its own missingness. Not-at-random (NAR) FCS is an appealing approach for conducting such sensitivity analyses \cite{leacy2017multiple,tompsett2018use}. Under this approach, the missingness indicator for the outcome is included as a predictor in the outcome imputation model, with the associated regression coefficient being a sensitivity parameter, referred to as ``delta’’ or $\delta$ (thus this sensitivity analysis approach is referred to as delta-adjustment). The value of the delta represents the difference between the expected value of the observed and missing outcomes conditional on other predictor variables in the imputation model. Therefore, the value of delta cannot be estimated from the observed data and must be specified using external information \cite{tompsett2020general}.

Despite the literature surrounding the importance of compatibility between the imputation and analysis models, compatibility is seldom discussed in the context of sensitivity analysis. Previous simulation studies to evaluate the performance of NARFCS only considered main-effects regression models as substantive analyses. In such settings, the ``naïve NARFCS’’, i.e. using the default main-effects regression models as univariate imputation models, is guaranteed to be compatible \cite{tompsett2018use}. However, if the naïve NARFCS is incompatible with the substantive analysis, the imputation may induce bias and render findings of sensitivity analyses unreliable. Addressing this issue is particularly important for estimating parameters such as the ACE, where the substantive analysis often goes beyond fitting a main-effects regression model, such as when using g-computation (also known as regression standardisation or the parametric g-formula) \cite{snowden2011implementation}. G-computation is a two-stage approach. In the first stage, an outcome regression model that would usually include exposure-confounder interaction terms is fitted. This model is then used in the second stage to estimate the average potential outcomes under two exposure levels. If interaction terms are included in the outcome model, the naïve NARFCS method would be incompatible with the substantive analysis as it does not incorporate these.

To address this gap, we propose two approaches for compatible sensitivity analysis when the outcome causes its own missingness in the context of multivariable missingness. The proposed approaches, NAR-SMCFCS and NAR-SMC-stack, extend MAR-SMCFCS and MAR-SMC-stack, respectively, to incorporate the delta-adjustment in the outcome imputation. We conduct a simulation study to evaluate the performance of the proposed approaches and compare them to the naïve NARFCS in a range of scenarios, focusing to fix ideas on the setting where the substantive analysis aims to estimate the ACE using g-computation. However, the proposed approaches apply more generally.

The paper is organised as follows. In section \ref{sec: Motivating example}, we introduce the motivating case study from the Victorian Adolescent Health Cohort Study (VAHCS) \cite{patton2002cannabis}. In section \ref{sec: Preliminaries and review of existing methods}, we first define the notation and the ACE estimand. We then review the outcome missingness mechanism and the NARFCS approach in section \ref{subsec: Missingness mechanisms and sensitivity analyses}, and the existing compatible imputation approaches (MAR-SMCFCS and MAR-SMC-stack) in section \ref{subsec: Compatibility and MAR-SMC approaches}. In section \ref{sec: Compatible sensitivity analysis}, we first introduce the proposed compatible sensitivity analysis approaches (NAR-SMCFCS and NAR-SMC-stack), and then detail the implementation of the proposed methods when using g-computation for ACE estimation in section \ref{subsec: Pooling ACE estimates after MI}. In section \ref{sec: Simulation study}, we describe the simulation study, based on the VAHCS case study. This is followed by a report of the application of the proposed methods to the VAHCS case study in section \ref{sec: Case study}. Finally, we summarise and discuss our findings in section \ref{sec: Discussion}.

\section{Motivating example}\label{sec: Motivating example}
The VAHCS is a prospective longitudinal cohort study ($n$=1943; $n$=1000 females) designed to investigate adolescent health and behaviour among participants recruited from schools in the state of Victoria, Australia, in 1992-1993 at ages 14 to 15 years old \cite{patton2002cannabis}. The study was approved by the Human Research Ethics Committee of the Royal Children’s Hospital, Melbourne, Australia. Participants were asked to complete a survey every six months in the first three years of the study (waves 2 to 6, adolescence phase) and later in 1998 at age 20 (wave 7, adulthood phase). An investigation using the VAHCS data by Patton et al. \cite{patton2002cannabis} examined the ACE of frequent cannabis use in female adolescents on the risk of mental health problems in young adulthood. We use this example as a case study, focusing on assessing the sensitivity of results to assumptions about missingness mechanisms under which the outcome causes its own missingness, which is plausible in this study given the nature of the outcome. 

Details of the case study measures are as follows. Participants were considered to be exposed if they reported use of cannabis more than once a week in any wave during adolescence (waves 2 to 6), and unexposed otherwise. We considered two outcome measures at wave 7, the mental health score and an indicator of mental health problems. The mental health score was a continuous outcome, derived by standardising and log-transforming the raw computerised Clinical Interview Schedule – Revised (CIS-R) score \cite{lewis1992manual}. The mental health problems indicator (binary outcome) was obtained by dichotomising the raw CIS-R score, with a score of 12 or higher indicating the presence of mental health problems and a score less than 12 indicating absence. The confounders, all of which were binary, consisted of two confounders measured at wave 2 (parental education and parental divorce or separation), and three confounders summarising measures across the adolescence phase (antisocial behaviour, adolescent depression and anxiety, and frequent alcohol use) \cite{patton2002cannabis}. For simplicity, thirty-nine females with missing values in three nearly complete confounders (parental education, parental divorce, and antisocial behaviour) were removed from the analysis. The analytical sample thus included $n$=961 participants. The participant's age at wave 2, log-transformed and standardised, was used as an auxiliary variable in MI. Frequent smoking in the adolescence phase was used to mimic an unmeasured common cause for generating correlated missingness indicators in the simulation study. Table \ref{tab: Table.one} presents the descriptive statistics and missing data proportions for each variable. 

\section{Preliminaries and review of existing methods}\label{sec: Preliminaries and review of existing methods}
\subsection{Notation}\label{subsec: Notation}
We consider a setting as in the VAHCS example, where the single time-point exposure, the outcome and a subset of the confounders are incomplete. We denote the exposure, outcome, set of complete and set of incomplete confounders by $X$, $Y$, $Z_1$, and $Z_2$, respectively, and the missingness indicator for each incomplete variable by $M$ with a subscript, e.g. $M_Y=1$ if $Y$ is missing and, $M_Y=0$ if $Y$ is observed. We assume that independent and identically distributed draws of ($X, Y, Z_1, Z_2$) and the missingness indicators are obtained from $n$ individuals. Let $f(Y|X, Z_1, Z_2, \theta)$ denote the model for the conditional distribution of the outcome given exposure and confounders used in the substantive analysis (see section \ref{subsec: Average causal effect and target analysis}), with corresponding parameter $\theta$. Throughout the paper, we refer to this model as the substantive outcome model and assume that it is correctly specified. To capture general cases in practice where unmeasured common causes of certain groups of variables exist, let $U$ denote the set of unmeasured common causes of exposure and confounders, and let $W$ denote the set of unmeasured common causes of missingness indicators. Additionally, let $V$ represent the set of all $p$ non-outcome variables that are incomplete in the dataset, that is, for which there is at least one missing value across the $n$ individuals, so that $\{X,Z_2\} \in V$, and let $V_{-j}=(V_1,\dots, V_{j-1},V_{j+1},\dots, V_p)$ denote the subset of $V$ by excluding $V_j$. Similarly, let $S$ represent the set of all $q$ non-outcome variables that are complete in the dataset, i.e. $Z_1 \in S$. In later sections when we consider auxiliary variables, $V$ and $S$ also include incomplete and complete auxiliary variables, respectively. 

\subsection{Average causal effect (ACE) and substantive analysis}\label{subsec: Average causal effect and target analysis}
Although the methods described and proposed in the paper apply more generally, to fix ideas we focus on a setting like VAHCS where the target estimand is the ACE, defined as the difference in the expected value (i.e. the mean or risk for a continuous or binary outcome, respectively) of the potential outcomes under exposure versus no exposure, where the potential outcome for an individual is the outcome that would have been observed if they were set to a particular exposure level (potentially counter to their actual exposure). In the absence of missing data, the ACE is non-parametrically identifiable from the observable data if the identifiability assumptions of exchangeability given confounders, consistency, and positivity hold \cite{Causal2020}. If this is the case, the ACE can be estimated using methods like g-computation. In g-computation, a regression model is first fitted for the conditional expectation of the outcome given the exposure and confounders, which we refer to as the substantive outcome model as mentioned above. This model usually includes exposure-confounder interaction terms which allows estimating the ACE under less restrictive parametric assumptions than the standard outcome regression approach. Specifically, it relaxes the assumption that the ACE is constant across confounder strata. The ACE is then estimated using the difference between the mean of outcome predictions from the substantive outcome model when setting all individuals to be exposed and when setting all individuals to be unexposed (regardless of their observed exposure) \cite{robins1986new,snowden2011implementation}. When the outcome is continuous, if there are no exposure-confounder interaction terms in the outcome model, g-computation reduces to the standard outcome regression approach, and the estimated regression coefficient for the exposure in the substantive outcome model provides an estimate of the ACE.

\subsection{Missingness mechanisms and sensitivity analyses}\label{subsec: Missingness mechanisms and sensitivity analyses}
\begin{figure}[h!]
\centering
\includegraphics[width=0.5\linewidth]{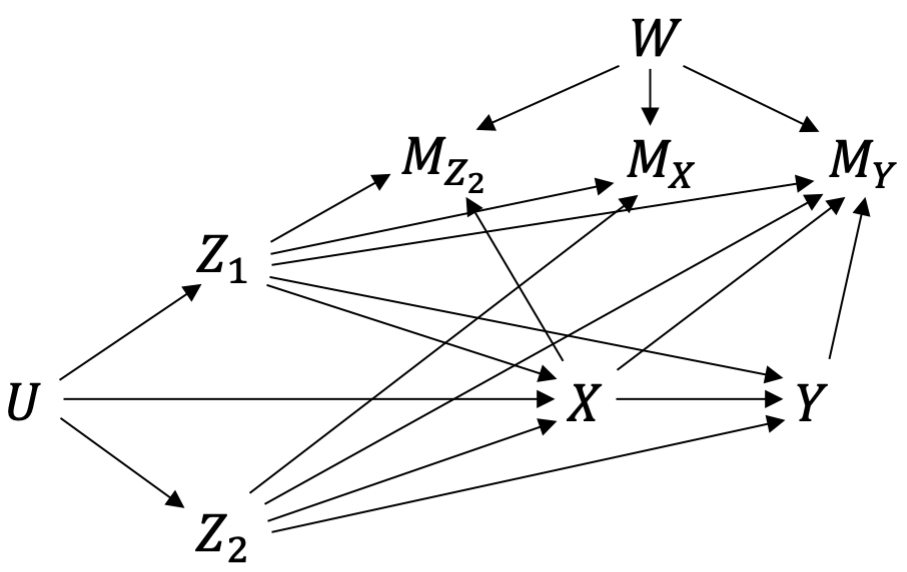}
\caption{Missingness directed acyclic graph (m-DAG) depicting a missingness mechanism where an incomplete outcome causes its own missingness. This is one of the canonical m-DAGs investigated in \cite{moreno2018canonical,zhang2024recoverability}.}
\label{fig: m-DAG}
\end{figure}

In the presence of missing data, if the target estimand is not non-parametrically identifiable from the observable data, it is said to be non-recoverable \cite{mohan2013graphical}. Previous research has shown that the ACE is non-recoverable in missingness mechanisms where the outcome causes its own missingness, which is a common occurrence in observational studies \cite{mohan2013graphical,mohan2014graphical,zhang2024recoverability}. An example of such a mechanism depicted by a missingness-directed acyclic graph (m-DAG) is presented in Figure \ref{fig: m-DAG}. 

In the rest of the manuscript, we focus on the m-DAG in Figure \ref{fig: m-DAG} to facilitate our examination of existing and proposed approaches, although in the discussion we note that the proposed approaches can be used under other missingness mechanisms. 

Previous studies have shown that if the arrow from the outcome to its missingness indicator is removed from the m-DAG in Figure \ref{fig: m-DAG}, the ACE is recoverable and can be unbiasedly estimated using for example compatible MI, without the need for sensitivity analysis \cite{zhang2024recoverability}. Thus, sensitivity analysis in this setting can be understood as assessing how the results change as the strength of the association between the outcome and its missingness indicator, induced by the arrow between them, varies. This can be addressed through a delta-adjustment approach, which is a parametric sensitivity analysis within a pattern-mixture framework \cite{rubin1977formalizing,mallinckrodt2014recent}. Under this approach, the association between $Y$ and $M_Y$ is captured by a sensitivity parameter delta ($\delta$), which represents the difference between the expected value of the missing and observed values, possibly conditional on other variables. For instance, in the simple setting of an incomplete continuous outcome, assuming just for this example that all other variables are complete, and that the substantive outcome model has no exposure-confounder interactions, the delta-adjustment approach might posit a model for the outcome as a function of exposure, confounders, and outcome missingness indicator as follows:
\begin{equation}\label{equ: out.delta}
E[Y|X,Z_1,Z_2,M_Y] = \alpha_0+ \alpha_1 X+ \alpha_2 Z_1 + \alpha_3 Z_2 + M_Y (\delta_0 + \delta_1 X).
\end{equation}

Here, $\delta_0$ and $\delta_0+\delta_1$ reflect the differences in mean outcome between the missing and observed outcomes in the unexposed and exposed groups, respectively, conditional on the confounders. The scenario where both $\delta_0$ and $\delta_1$ are zero corresponds to a lack of association between $Y$ and $M_Y$ within exposure groups. The scenario where $\delta_1=0$ corresponds to the setting where the difference between the conditional means of missing and observed outcomes is the same for exposed and unexposed groups. 

In the multivariable missingness setting, which is our focus in this manuscript, the delta-adjustment for outcome missingness can be implemented in MI using the NARFCS procedure by specifying a model like (\ref{equ: out.delta}) as the univariate imputation model for the outcome \cite{leacy2017multiple,tompsett2018use,tompsett2020general}. As in standard FCS, the outcome and other variables are imputed sequentially using a chained equations algorithm. Under the m-DAG in Figure \ref{fig: m-DAG}, because there are no arrows between non-outcome variables and their own missingness, these variables do not need the delta-adjustment and are imputed using univariate imputation models as in the standard implementation of the FCS procedure, i.e. without including missingness indicators. An imputation model like (\ref{equ: out.delta}) consists of two components, which may be termed the identifiable and non-identifiable parts. The identifiable part consists of terms involving other variables, i.e. $\alpha_0+ \alpha_1 X+ \alpha_2 Z_1 + \alpha_3 Z_2 $. The non-identifiable part includes the missingness indicator for the variable to be imputed, here the outcome, i.e. $M_Y(\delta_0+\delta_1 X)$. Within each iteration of the chained equations algorithm of NARFCS, coefficients in the identifiable part, i.e. $\alpha$’s, are obtained by fitting the model to the records with the complete outcome, with other incomplete variables set to their imputed values under the current iteration. For the non-identifiable part, the $\delta$s are fixed throughout the NARFCS procedure to values elicited from external information such as experts’ knowledge \cite{leacy2017multiple,tompsett2018use,tompsett2020general}. It is recommended to specify a range of plausible values for the $\delta$s, and repeat the NARFCS procedure under each plausible value. 

An outcome imputation model like (\ref{equ: out.delta}) is incompatible with a substantive model that excludes $M_Y$. This is a common feature of delta-adjustment approaches like NARFCS and has been found to be a kind of incompatibility that does not result in important biases \cite{liu2014stationary,leacy2017multiple}. Meanwhile, when the substantive outcome model has exposure-confounder interactions, an inappropriate implementation of the NARFCS approach, for example based on univariate imputation models with no interaction terms in the identifiable part, as is common, is a kind of incompatibility that may lead to biases of more concern. In this case, using a delta-adjusted version of the substantive outcome model (with its interactions) to impute the outcome is not sufficient to ensure compatibility in the presence of multivariable missingness; outcome-covariate interactions also need to be included as predictors when imputing other variables \cite{tilling2016appropriate}. Additionally, manually adjusting imputation models cannot eliminate other forms of incompatibility such as those resulting from non-linear terms, e.g. quadratic terms, or non-linear substantive outcome models, e.g. Cox proportional hazard model. The next section introduces two compatible MI approaches that are more flexible for handling general forms of incompatibility of this kind.

\subsection{Compatibility and MAR-SMC approaches in the complete outcome setting}\label{subsec: Compatibility and MAR-SMC approaches}
For brevity, this subsection reviews existing approaches by focussing on the complete outcome setting. The imputation model is said to be compatible with the substantive model if there exists a joint model such that the imputation model and the substantive outcome model are the corresponding conditional models of the joint model \cite{bartlett2015multiple,liu2014stationary}. For instance, suppose a joint model for the joint distribution of incomplete non-outcome variables and outcome is denoted by $f(X,Y,Z_2|Z_1,\theta, \lambda)$, which has the substantive outcome model $f(Y|X,Z_1,Z_2, \theta)$ as its corresponding conditional model. Then, a compatible imputation model for the exposure, for example, should be the corresponding conditional model of this joint distribution model, i.e., the imputation model is given by $f(X|Z_1,Z_2,Y, \lambda, \theta)$, which is proportional to $f(Y|X,Z_1,Z_2,\theta)f(X|Z_1,Z_2,\lambda)$. Therefore, imputations for the exposure drawn in a way that is proportional to this latter product of distributions are considered compatible. In the general case with multiple incomplete non-outcome variables (with this variable set being denoted by $V$, per section \ref{subsec: Notation}), using the terminology of importance sampling literature, we refer to $f(Y|X,Z_1,Z_2,\theta)f(V_j|V_{-j},S,\lambda_j)$ as the target distribution for compatible imputation of $V_j$ and $f(V_j|V_{-j},S,\lambda_j)$ as the corresponding proposal distribution. Below, we review the two MAR-SMC approaches for drawing imputations from this target distribution in the complete outcome setting.

\textbf{The MAR-SMCFCS approach} draws values for $V_j$ from the corresponding proposal distribution, i.e. $V_j^* \sim f(V_j|V_{-j},S,\lambda_j)$, and then imputations for $V_j$ are drawn from the target distribution evaluated at the sampled value $V_j^*$, which is obtained by first evaluating the substantive outcome model at that value. The procedure to draw imputations can be done directly for binary and categorical variables, while rejection sampling or other sampling methods can be used for continuous or count variables \cite{bartlett2015multiple}. Within each iteration of the chained equations algorithm, the sampling process is repeated to update the imputed values for each incomplete non-outcome variable (if more than one) sequentially. The iteration is repeated until convergence, and a final imputed dataset is obtained. This process is repeated multiple times to obtain multiple imputed datasets. Finally, the substantive analysis is conducted within each imputed dataset, and the final results are obtained using Rubin's rules \cite{rubin2004multiple}.

\textbf{The MAR-SMC-stack approach} uses FCS to impute each non-outcome variable $V_j$ from its corresponding proposal distribution, $f(V_j|V_{-j},S,\lambda_j)$. Then, using importance sampling \cite{meng1994multiple}, the imputations are weighed by the substantive outcome model distribution $f(Y|X,Z_1,Z_2,\theta)$ evaluated at the observed $Y$ and $Z_1$ and the imputed values for $X$ and $Z_2$, which is equal to the ratio between the target distribution and the proposal distribution -- also known as the importance ratio \cite{beesley2021stacked}. The aim of the weighting is to correct the imputations of non-outcome variables for the fact that they may not be compatible with the substantive outcome model. Specifically, the MAR-SMC-stack approach imputes incomplete non-outcome variables sequentially using the FCS procedure from the dataset with the outcome removed and obtains $M$ imputed datasets. Next the outcome is added back into each imputed dataset, and these are then stacked on top of each other, with each row in the stacked dataset assigned a weight proportional to the substantive outcome model distribution. Lastly, the substantive outcome model is fitted on the stacked dataset using the assigned weights, to obtain the point estimate and its estimated variance, denoted by $\widehat{\theta}$ and $\widehat{Var}_{stack}(\widehat{\theta})$ respectively. Wood et al. showed that, with maximum likelihood estimation, the point estimate $\widehat{\theta}$ approximately equals the point estimate given by Rubin's rules and the estimated variance $\widehat{Var}_{stack}(\widehat{\theta})$ is approximately equal to $1/M$ times the average of the within-imputation variance \cite{wood2008should}. The following formula proposed by Beesley and Taylor gives the overall variance for the point estimator from the stacked approach \cite{beesley2021accounting}, which we call Beesley’s rule. 
\begin{equation}\label{equ: Beesley’s rule}
\widehat{Var}(\widehat{\theta})= M\widehat{Var}_{stack}(\widehat{\theta}) + (M+1)\widehat{Var}_{between}(\widehat{\theta}),
\end{equation}
where $\widehat{Var}_{between}(\widehat{\theta})$ is the estimated between-stack imputation variance, which can be obtained using a jackknife estimator \cite{beesley2021accounting}.

\section{Compatible sensitivity analysis }\label{sec: Compatible sensitivity analysis}
In this section, we come back to our original setting of interest with multivariable missingness, including missingness in the outcome. Both MAR-SMCFCS and MAR-SMC-stack ensure compatibility, which provides the rationale for proposing extensions to these approaches for conducting compatible sensitivity analysis under an assumed missingness mechanism where the outcome may cause its own missingness (Figure \ref{fig: m-DAG}). In the proposed approaches, described next, the outcome imputation model is an extension of the substantive outcome model incorporating a delta-adjustment, as in model (\ref{equ: out.delta}), i.e. including a non-identifiable part involving the outcome missingness indicator and associated sensitivity parameters ($\delta$s). In such cases, we denote the outcome imputation model as $f(Y|X, Z_1, Z_2, M_Y, \theta^\prime, \delta)$, where $\theta^\prime$ denotes the parameter in the identifiable part -- it is different to $\theta$ as this model is conditional on $M_Y$. Of note, this outcome imputation model (with non-zero $\delta$s) is incompatible with the substantive outcome model as the latter does not involve the missingness indicator $M_Y$ \cite{liu2014stationary,leacy2017multiple}. As previously mentioned, this kind of incompatibility is not expected to be problematic based on the previous research \cite{leacy2017multiple,tompsett2018use}. The following proposed approaches were developed to avoid other kinds of incompatibility that may result in biases of more concern.

\subsection{NAR-SMCFCS approach}\label{sec: NAR-SMCFCS approach}
Here we describe the proposed extension to MAR-SMCFCS, referred to as the not-at-random SMCFCS approach, or NAR-SMCFCS. Algorithm \ref{nar-smcfcs} details the steps involved in the approach, where steps 2 and 8 reflect the needed modifications to MAR-SMCFCS. Within each iteration of the chained equations algorithm, first all non-outcome variables are imputed, then the NAR-SMCFCS approach imputes the outcome from the delta-adjusted model $f(Y|X, Z_1, Z_2, M_Y, \theta^\prime, \delta)$ using the latest imputed values for other variables.

\begin{breakablealgorithm}
\caption{NAR-SMCFCS}\label{nar-smcfcs}
\begin{enumerate}
\item Draw initial imputations for each incomplete non-outcome variable $V_j$ by performing random draws with replacement from observed records.
\item Draw initial imputations for the outcome by sampling from the outcome imputation model $f(Y|X, Z_1, Z_2, M_Y, \theta^\prime, \delta)$, where the parameters in the identifiable part $\theta^\prime$ are estimated by fitting the substantive outcome model to the records with the outcome observed (with initial imputed values for other variables), and parameters in the non-identifiable part (i.e. the $\delta$s) are fixed to elicited values. 
\item For iteration $t=1,\dots,T$, repeat steps 4 to 8.
\item For non-outcome variables $j=1,\dots,p$, repeat steps 5 to 7.
\item Fit the substantive outcome model $f(Y|X, Z_1, Z_2, \theta)$ to the current imputed dataset. 
\item Fit the proposal distribution for the incomplete non-outcome variable $V_j$, $f(V_j|V_{-j},S,\lambda_j)$, to the current imputed dataset.  
\item Impute missing values for the incomplete non-outcome variable $V_j$ by drawing from the target distribution $f(Y|X,Z_1,Z_2,\theta)f(V_j|V_{-j},S,\lambda_j)$, either directly or using rejection sampling as in the MAR-SMCFCS procedure, based on the models fitted in steps 5 and 6. 
\item Impute the missing values in the outcome by sampling from the outcome imputation model $f(Y|X, Z_1, Z_2, M_Y, \theta^\prime, \delta)$, where the parameters in the identifiable part $\theta^\prime$ are estimated by fitting the substantive outcome model to the records with the observed outcome in the current imputed dataset, and the parameters in the non-identifiable part (i.e. the $\delta$s) are fixed to elicited values. 
\item This yields one imputed dataset used as one of the $M$ imputed datasets.
\end{enumerate}
\end{breakablealgorithm}

As in MAR-SMCFCS, to impute each partially observed non-outcome variable $V_j$, the proposal distribution is $f(V_j|V_{-j},S,\lambda_j)$. The imputation for $V_j$ is thus made compatible with the substantive outcome model by sampling (directly or using rejection sampling) from the target distribution $f(Y|X,Z_1,Z_2,\theta)f(V_j|V_{-j},S,\lambda_j)$, where at each iteration the parameter $\theta$ is estimated from the current imputed dataset using the imputation for the outcome obtained at the previous iteration using the delta-adjusted model. 

\subsection{NAR-SMC-stack approach}\label{subsec: NAR-SMC-stack approach}
We refer to the extension of the MAR-SMC-stack approach for sensitivity analysis as NAR-SMC-stack. Algorithm \ref{nar-smc-stack} details the steps involved in the approach. The NAR-SMC-stack approach starts by imputing non-outcome variables multiple times in FCS, then stacks the imputed datasets and uses the stacked dataset to impute the outcome. 

\begin{breakablealgorithm}
\caption{NAR-SMC-stack}\label{nar-smc-stack}
\begin{enumerate}
\item Obtain $M$ imputations for non-outcome variables $V_j$, $j=1,\dots,p$ using a standard FCS-based approach, with univariate imputation models $f(V_j|V_{-j}, S, M_Y, \lambda_j)$ including the outcome missingness indicator as predictor but excluding the outcome. 
\item Stack the $M$ imputed datasets.
\item Impute the missing values in the outcome by sampling from the outcome imputation model $f(Y|X, Z_1, Z_2, M_Y, \theta^\prime, \delta)$ evaluated at the imputed values of non-outcome variables, where the parameters in the identifiable part $\theta^\prime$ are estimated by fitting the outcome imputation model to fully observed records in the original dataset, and parameters in the non-identifiable part (i.e. the $\delta$s) are fixed to elicited values. 
\item Assign a weight for each record in the stacked dataset, with weights as specified in equation (\ref{equ: nar-smc-stack.wt}), where the parameters required for obtaining the weights are estimated in step 3.
\end{enumerate}
\end{breakablealgorithm}

In step 1, the outcome missingness indicator $M_Y$ is regarded as a fully observed covariate and included as a predictor in the univariate imputation models for non-outcome variables. This is done to ensure compatibility with the outcome imputation model. Indeed, when the outcome imputation is delta-adjusted, the implied difference in the outcome distribution between the observed and missing values should be reflected when imputing the non-outcome variables. Specifically, the joint distribution for compatible imputation is given by $f(V,Y|S,M_Y,\lambda,\theta^\prime, \delta)$, which has the corresponding conditional distribution $f(Y|X,Z_1,Z_2,M_Y,\theta^\prime, \delta)$ for imputing the outcome and the corresponding conditional distribution $f(V_j|V_{-j},S,Y,M_Y,\lambda)$ for compatibly imputing the non-outcome variable $V_j$. 

For NAR-SMC-stack, we compute the weights for step 4 based on the importance ratio for each missingness pattern:
\begin{itemize}
\item Missingness pattern I: individual has no missing values. No imputation is needed, and the importance ratio is 1. 
\item Missingness pattern II: individual has no missing values in any of the incomplete non-outcome variables (i.e. the variables in $V$) but has missingness in the outcome. The outcome is therefore imputed from the outcome imputation model evaluated at the observed values for the other variables. There are no imputations of non-outcome variables to weigh, and the importance ratio is 1. 
\item Missingness pattern III: individual has outcome value observed but missing values in the non-outcome variables $V$. The conditional distribution for imputing $V_j$ compatibly, i.e. the target distribution, is $f(V_j|V_{-j}, S, Y, M_Y=0, \lambda_j,\theta^\prime)$ and the proposal distribution is $f(V_j|V_{-j}, S, M_Y=0, \lambda_j)$. Thus, the importance ratio, computed from the ratio between the target distribution and the proposal distribution, is given by: \begin{equation*}
\begin{aligned}
\frac{f(V_j|V_{-j},S,Y, M_Y=0, \lambda_j,\theta^\prime)} {f(V_j|V_{-j},S,M_Y=0, \lambda_j)} &\propto f(Y|V,S,M_Y=0,\theta^\prime)\\&=f(Y|X, Z_1,Z_2, M_Y=0, \theta^\prime).
\end{aligned} 
\end{equation*}
\item Missingness pattern IV: individual has missingness in both outcome and non-outcome variables. As above, the joint distribution for compatible imputation is $f(V,Y|S,M_Y=1,\lambda, \theta^\prime, \delta)$, which has as decomposition $f(V|S,M_Y=1, \lambda)f(Y|X,Z_1,Z_2,M_Y=1, \theta^\prime, \delta)$. Since in the NAR-SMC-stack approach, variables in $V$ are imputed in the FCS algorithm from $f(V|S,M_Y=1, \lambda)$, and then the outcome is imputed from $f(Y|X,Z_1,Z_2,M_Y=1, \theta^\prime, \delta)$ using the imputations of the variables in $V$, the imputations are automatically compatible and no weighting is needed. Here also, the importance ratio is 1. 
\end{itemize}
The importance ratio in missingness pattern III is computed using the estimated conditional outcome distribution under the outcome imputation model fitted to the complete cases \cite{beesley2021stacked}. This estimated distribution is evaluated for each individual $i$ at the observed outcome $Y_i$ given the imputed values of $X_i$ and ${Z_2}_i$, denoted by $X_i^m$ and ${Z_2}_i^m$. For instance, if the outcome imputation model is a normal linear regression with a linear predictor given by $h(X,Z_1,Z_2,M_Y=0,\theta^\prime)$, the importance ratio for individual $i$ is computed as follows: 
$$f(Y_i|X_i^m,{Z_1}_i,{Z_2}_i^m,{M_Y}_i=0,\widehat{\theta^\prime},\hat{\sigma}_{\epsilon}^2)=\frac{\exp (-\frac{(Y_i-h(X_i^m,{Z_1}_i,{Z_2}_i^m,{M_Y}_i=0,\widehat{\theta^\prime}))^2} {2\hat{\sigma}_{\epsilon}^2})}{\sqrt{2\pi \hat{\sigma}_{\epsilon}^2}},$$
where $\widehat{\theta^\prime}$ and $\hat{\sigma}_{\epsilon}^2$ are estimated by fitting the outcome imputation model to the complete cases.

Therefore, the desired weights assigned to each record in the stacked dataset in step 4 are computed from the importance ratios for each missingness pattern as described above. Once all of the weights have been calculated, they are rescaled so that the sum of the weights within the individual is 1. Thus, the weights for individual $i$ and imputation $m$ are given by:
\begin{equation} \label{equ: nar-smc-stack.wt}
w_{i,m}=\begin{cases}
\frac{f(Y_i|X_i^m,{Z_1}_i,{Z_2}_i^m,{M_Y}_i=0,\widehat{\theta^\prime})} {\sum_{m=1}^M f(Y_{i}| X_i^m,{Z_1}_i,{Z_2}_i^m,{M_Y}_i=0,\widehat{\theta^\prime})} & \text{, for individual $i$ in missingness pattern III}\\
1/M & \text{, else}
\end{cases}.
\end{equation}

\subsection{Pooling ACE estimates after MI}\label{subsec: Pooling ACE estimates after MI}
When the substantive analysis consists of fitting a regression model, and the target estimand is a regression coefficient, the final MI estimate and corresponding standard error from NAR-SMCFCS or NAR-SMC-stack can be obtained using Rubin’s rules or Beesley’s rule, respectively. However, when the substantive analysis is g-computation, additional considerations are needed when pooling the results because the standard error in g-computation is generally estimated using the bootstrap. This section provides details for pooling the g-computation estimates of the ACE after imputation with the approaches proposed in this paper. The strategies are applicable in broader settings whenever the substantive analysis uses bootstrapping for estimating standard errors.

With the NAR-SMCFCS algorithm, after creating $M$ imputed datasets, the substantive analysis is applied to each imputed dataset. Specifically, within each imputed dataset, g-computation is used to obtain a point estimate of the ACE and bootstrap is used to obtain a corresponding variance estimate. The resulting $M$ point and variance estimates can then be pooled using Rubin’s rules. This approach has been shown to achieve approximately nominal coverage for compatible imputation by Bartlett and Hughes \cite{bartlett2020bootstrap}. The simulation study \ref{subsec: Missing data approaches} implemented this strategy, referred to as ``MI bootstrap Rubin’’.  

With the NAR-SMC-stack approach, when implementing g-computation, the substantive outcome model is fitted with weighting to the stacked dataset using the assigned weights. The fitted model is then used to predict outcomes for all records and compute their mean when the exposure is set to the values to be contrasted (e.g. 1 and 0 in the case that the exposure is coded 0/1). The difference between these means provides an estimate of the ACE. The bootstrap is used to obtain the within-stack imputation variance, $\widehat{Var}_{stack}(\widehat{ACE})$. The between-stack imputation variance estimate $\widehat{Var}_{between}(\widehat{ACE})$ can be obtained by the jackknife estimator. Specifically, each jackknife dataset comprises $M-1$ imputed datasets, and the weights are re-scaled over $M-1$ imputations for each individual to sum up to 1. G-computation is applied to each jackknife dataset to estimate the ACE (using a weighted outcome model as described above), from which the jackknife variance estimate is obtained. 

\section{Simulation study}\label{sec: Simulation study}
To compare the performance of the methods described in Section \ref{sec: Compatible sensitivity analysis}, we designed a simulation study based on the VAHCS case study, with data generated following the m-DAG in Figure \ref{fig: m-DAG} but with the unmeasured $U$ replaced by the measured auxiliary variable $A$ (age at wave 2 in VAHCS as shown in Table \ref{tab: Table.one}, but assumed complete for the simulations). The set of complete confounders $Z_1$ consisted of $C_1,C_2,C_3$, the set of incomplete confounders $Z_2$ consisted of $C_4$ and $C_5$ as defined in Table \ref{tab: Table.one}, and adolescent smoking played the role of the unmeasured variable $W$ as shown in that table.

\subsection{Data generation}
The parameter values used for the data generation models were estimated by fitting analogous models to the VAHCS data unless stated otherwise. Values are provided in the Supplementary Material. We started by separately generating one continuous auxiliary variable $A$ and one binary confounder $C_1$ from standard normal and Bernoulli distributions, respectively. Then we generated the rest of the confounders ($C_2$ to $C_5$) and the exposure from Bernoulli distributions in sequence. For each variable, the success probability was given by a logistic regression model conditioning on all the previously generated variables. The exposure prevalence was controlled to be higher than in VAHCS, at approximately 50\%, by modifying the model intercept. The outcome was generated using a generalised linear model with mean specification: 
\begin{equation}
\begin{aligned}\label{equ: out.gen}
    g({\rm E}(Y|C, X))&=\beta_0+\beta_1 C_1+\beta_2 C_2+\beta_3 C_3+\beta_4 C_4+\beta_5 C_5+\beta_6 X+\beta_7 XC_4\\
&+ \beta_{1,4} C_1C_4+\beta_{2,4} C_2C_4+ \beta_{3,4} C_3C_4+\beta_{4,5} C_4C_5+\beta_{3,5} C_3C_5.
\end{aligned}
\end{equation} 
The link function $g(\cdot)$ was the identity for continuous, and logit for binary outcome scenarios. The continuous outcome was generated from a Gaussian distribution (with standard deviation set to 1), and the binary outcome from a Bernoulli distribution. We considered three outcome generation scenarios: no/weak/strong $XC_4$ interaction, by setting the coefficient for the exposure-confounder interaction term $\beta_7$ to 0, $-0.5$, and $-3$ times the main effect $\beta_6$, respectively. We modified the strength of the main effect $\beta_6$ to fix the true value of the target parameter, the ACE, to be a 0.3 difference in means for the continuous outcome, and a 0.12 risk difference for the binary outcome, across all scenarios. We generated samples of 1000 records, which achieved approximately 80\% power for the conventional 0.05 significance level in all scenarios (before introducing any missing values) for the specified ACE values. In most scenarios the target analysis for estimating the ACE was g-computation with the substantive outcome model specified as the outcome generation model (\ref{equ: out.gen}) for each scenario, and the standard error (SE) was obtained by bootstrapping. The only exception was the no-interaction scenario ($\beta_7=0$) with the continuous outcome, in which, instead of g-computation, we used outcome regression as the substantive analysis, i.e. we fit a linear regression model with no exposure-confounder interactions and used the estimate of $\beta_6$ and its conventionally estimated model-based SE as the estimates of the ACE and its SE.

\subsection{Missingness generation}
Again mimicking the VAHCS data, we imposed missingness on two confounders ($C_4$ and $C_5$), the exposure, and the outcome, with the missingness proportions controlled at 15\%, 15\%, 20\%, and 20\%, respectively. The overall complete-case proportions were around 55\% and 50\% for the continuous and binary outcome settings, respectively. To do this, we first generated the binary variable $W$, and used it as an unmeasured common cause for all missingness indicators as depicted in Figure \ref{fig: m-DAG}. The missingness indicator for each incomplete confounder was generated from a logistic regression model conditional on $W$, exposure, and the complete confounders ($C_1$ to $C_3$). The generation model for the exposure missingness indicator used $W$ and all confounders ($C_1$ to $C_5$) as predictors. We considered two scenarios for generating the missingness in the outcome using logistic regression. The simple missingness scenario generated the outcome missingness indicator using a main effects model with confounders, exposure, and outcome as covariates, where the regression coefficient for the outcome was fixed at log(3). In the complex missingness scenario, the model included additionally an exposure-outcome interaction term, with the regression coefficient fixed at log(2). Therefore, in the complex scenario, the probability of the outcome being missing differed between exposure groups (for the same outcome value). 

\subsection{Missing data approaches}\label{subsec: Missing data approaches}
We evaluated the two proposed approaches and compared them to a naïve implementation of the NARFCS approach, described below. In all approaches, the univariate imputation model for each non-outcome variable was a logistic regression, with the other non-outcome analysis variables and the auxiliary variable $A$ as predictors. The imputation models for non-outcome variables in the naïve NARFCS approach used main-effects univariate imputation models (i.e. without interaction terms) that also included the outcome as a predictor, while the imputation models in the NAR-SMC-stack approach instead included the outcome missingness indicator as a predictor (except in scenarios where the sensitivity parameters were set to null). The outcome imputation model was the same for all 3 approaches, including the naïve NARFCS, so that all approaches were evaluated with the same values for the sensitivity parameters. The outcome imputation model was as below (\ref{equ: out.imp}) for the complex missingness and weak/strong interaction outcome scenarios. The non-identifiable part of the model only included $\delta_0 M_Y$ in the simple missingness scenario. The term $\beta_7 XC_4$ was removed from the model (\ref{equ: out.imp}) in the no-interaction outcome scenarios. 
\begin{equation}
\begin{aligned}\label{equ: out.imp}
    g({\rm E}(Y|X,C,M_Y))&=\beta_0+\beta_1 C_1+\beta_2 C_2+\beta_3 C_3+\beta_4 C_4+\beta_5 C_5+\beta_6 X+\beta_7 XC_4\\
    &+ \beta_{1,4} C_1C_4+\beta_{2,4} C_2C_4+ \beta_{3,4} C_3C_4+\beta_{4,5} C_4C_5+\beta_{3,5} C_3C_5\\
&+\delta_0 M_Y + \delta_1 M_Y X.
\end{aligned}
\end{equation} 
Approximate true values for the $\delta$s were estimated by fitting the (potentially misspecified) outcome imputation model (\ref{equ: out.imp}) to a large generated complete dataset (without missing data). See Discussion. These values are available in the Supplementary Material.

To investigate the performance of the missing data approaches over a range of sensitivity parameters, we implemented each method in three ways: by setting $\delta_0$ to the true value, to twice the true value, and to zero with each approach. For each approach, the chained equation algorithm was iterated five times, and we generated 20 multiply imputed datasets. For NAR-SMCFCS and naïve NARFCS, ACE estimates from g-computation and bootstrapped SEs (over 200 bootstrapped samples) across the 20 imputed datasets were pooled by applying Rubin’s rules. For the NAR-SMC-stack, the point and variance estimates were obtained as described in section \ref{subsec: Pooling ACE estimates after MI}. 

We generated 2000 datasets for each scenario and obtained the following performance indicators: the mean of the ACE estimates (recalling that these are in the mean difference scale for the continuous outcome and the risk difference scale for the binary outcome), the relative bias (RB) compared to the true value (as a percent, \%), the empirical SE of the estimated ACE, the average of the estimated SE, the mean square error, the 95\% confidence interval (CI) coverage probability, and the Monte Carlo SE for all measures; see the Supplementary Material for formulae. All analyses were carried out in R version [4.1.2] \cite{Rversion}. 

\subsection{Simulation results}
\begin{figure}[h!]
\centering
\includegraphics[width=0.9\linewidth]{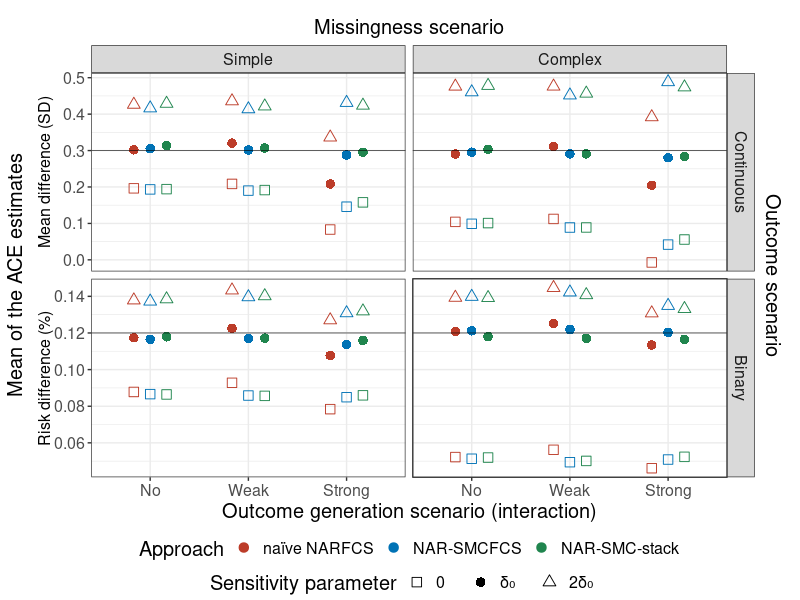}
\caption{Simulation study results: Mean ACE estimates for the three missing data methods for a range of sensitivity parameter values, set to 0 (hollow squares), the true value, $\delta_0$ (solid circle), and twice the true value, $2\delta_0$ (hollow triangle), for continuous (top panel) and binary (bottom panel) outcome settings, across three outcome generation and two outcome missingness generation scenarios.}
\label{fig: simu.res}
\end{figure}

Figure \ref{fig: simu.res} (top panel) shows the mean ACE estimates across the simulated datasets for the continuous outcome setting for the different outcome generation and missingness scenarios. When the outcome generation model had no exposure-confounder interactions, incompatibility was not an issue, and accordingly, the three approaches showed similar results. In this setting, the ACE estimates were approximately unbiased (RBs $\leq \pm 4\%$) for all three approaches when the sensitivity parameter was set to the true value. This is consistent with our expectation and the findings from previous literature that the naïve NARFCS works well in this setting (when the outcome causes its own missingness) if the correct external information is used. 

However, in the scenarios with an exposure-confounder interaction, the naïve NARFCS approach was incompatible with the analysis model, which led to bias in the estimation of the ACE even when the sensitivity parameters were set to their true values, with larger bias observed as the interaction term strengthened. Specifically, the mean ACE estimates from the naïve NARFCS in the simple and complex missingness scenarios were 0.32 (6.8\% RB) and 0.31 (3.5\% RB), respectively, in the weak interaction scenario, and 0.21 (-30.6\% RB) and 0.20 (-31.8\% RB), respectively, in the strong interaction scenario. In comparison, the two proposed approaches were approximately unbiased in these scenarios (the RB was less than $\pm 3\%$ in the weak interaction scenario and less than $\pm 6\%$ in the strong interaction scenario for both approaches). In the strong interaction scenario, the mean ACE estimates given by the naïve NARFCS were shifted downwards compared to the two proposed approaches when the sensitivity parameter was set to the incorrect value. For instance, the mean ACE estimates from the naïve NARFCS were 0.08 and 0.34 in the scenario with strong interaction and simple missingness when the sensitivity parameter was set to zero and twice the true $\delta_0$, respectively. In contrast, the estimates given by NAR-SMCFCS were 0.15 and 0.43 (for zero and twice the true $\delta_0$, respectively) in those scenarios, representing an approximately symmetric range around the true ACE (0.3), with similar estimates from the NAR-SMC-stack approach. 

Figure \ref{fig: simu.res} (bottom panel) shows the mean ACE estimates across the scenarios for the binary outcome setting. Although the risk difference is not directly comparable with the mean difference, the results for the binary outcome were consistent with those observed in the continuous outcome setting in terms of the general patterns. The proposed NAR-SMC approaches were approximately unbiased for all scenarios. In contrast, the naïve NARFCS were somewhat biased when the outcome generation model included an exposure-confounder interaction, i.e. the mean ACE estimates (risk differences, expressed as a percent) in the strong interaction outcome scenario were 10.8\% (10\% RB) and 11.3\% (5.8\% RB) for the simple and complex missingness scenarios, respectively. Note that in this setting, we do not expect symmetry over the range of sensitivity parameters, as the $\delta$s are on a log scale.

\begin{figure}[h!]
\centering
\includegraphics[width=\linewidth]{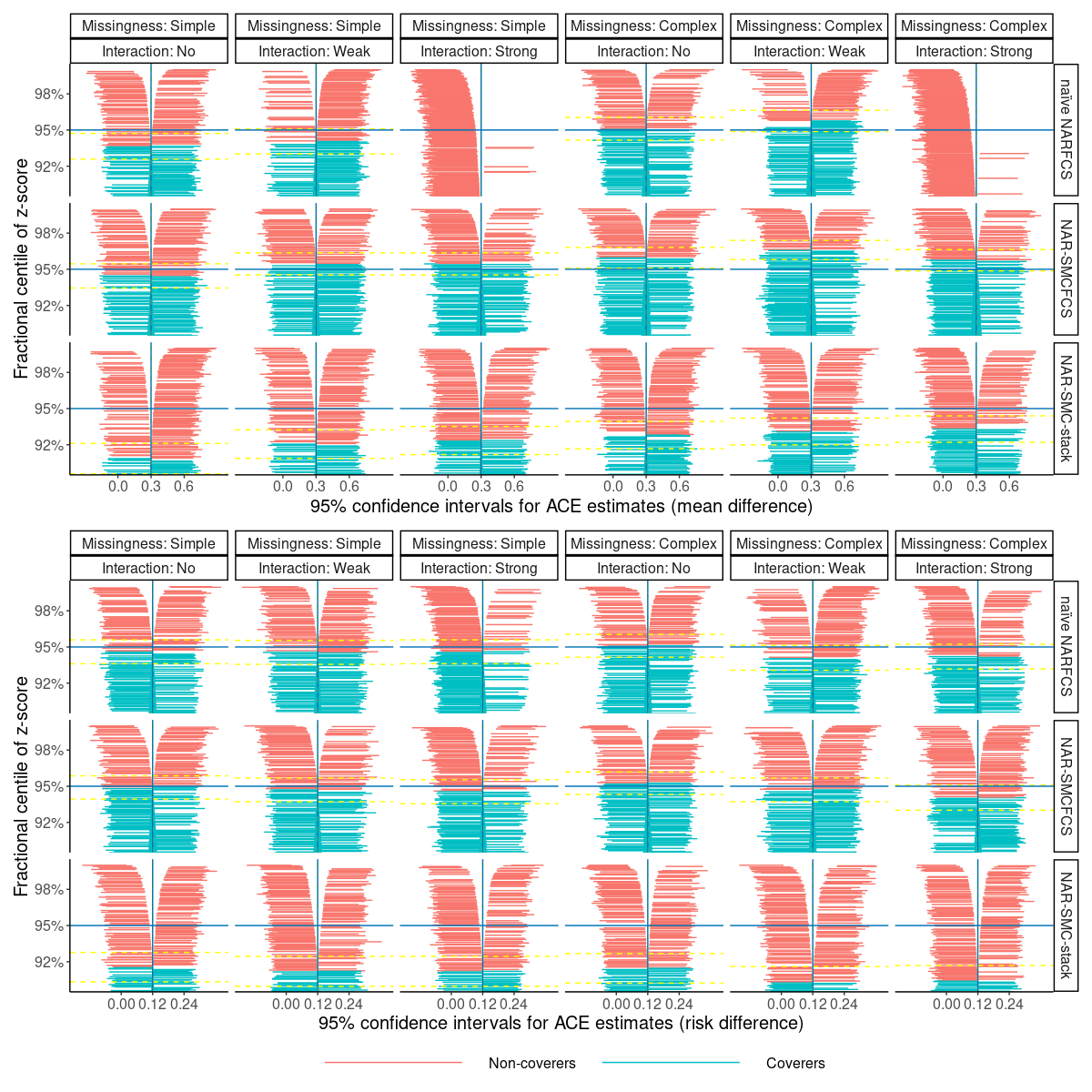}
\caption{Simulation study results: Zipper plot for the three missing data methods using the true value of sensitivity parameters, for continuous (top panel) and binary (bottom panel) outcome settings, across three outcome generation and two outcome missingness generation scenarios.}
\label{fig: Simu.zipper}
\end{figure}

Table \ref{tab: Simu.coverage} and Figure \ref{fig: Simu.zipper} present, respectively, the estimated coverage probabilities of the 95\% CI and the zipper plots \cite{morris2019using} when the true sensitivity parameters were used in the missing data approaches. The NAR-SMCFCS (all scenarios) and naïve NARFCS (in the no or weak interaction outcome scenarios) achieved approximately 95\% coverage. In the strong interaction outcome scenarios, the coverage for the naïve NARFCS was below nominal with continuous but not binary outcome. Therefore, the magnitude of bias in the binary outcome settings considered was not large enough to have consequences on coverage, but it could have larger impact in scenarios with stronger interactions. The coverage given by the NAR-SMC-stack approach was slightly below the nominal value (around 92\% across all scenarios). The model-based SE given by NAR-SMC-stack was also smaller compared to the NAR-SMCFCS approach, i.e., SEs ranged from 0.009 to 0.103 for NAR-SMC-stack and from 0.108 to 0.112 for NAR-SMCFCS in the continuous outcome setting, and from 0.040 to 0.043 for NAR-SMC-stack versus 0.044 to 0.047 for NAR-SMCFCS in the binary outcome setting. The coverage and model-based SEs suggest a potential problem with variance estimation in the NAR-SMC-stack.

\section{Application to the VAHCS case study}\label{sec: Case study}
As noted in section \ref{sec: Motivating example}, the missingness in the outcome (mental health score) in the VAHCS analysis was likely to be caused by the outcome \cite{cheung2017impact}, and the m-DAG in Figure \ref{fig: m-DAG} was deemed to depict a plausible missingness mechanism for the motivating example. Therefore, in this section, we conducted a sensitivity analysis to assess the robustness of the results across a range of plausible outcome missingness assumptions. We focused on the continuous outcome measurement, i.e., the mental health score. The substantive analysis consisted of g-computation to estimate the ACE of cannabis use in adolescence on mental health score in young adulthood among females using the substantive outcome model specified in (\ref{equ: out.case}): \begin{equation}\label{equ: out.case}
{\rm E}(Y|X,C)=\eta_0+\eta_1 C_1+\eta_2 C_2+\eta_3 C_3+\eta_4 C_4+\eta_5 C_5+\eta_6 X+\eta_7 XC_3.
\end{equation} 
This model incorporates an interaction between exposure and antisocial behaviour ($C_3$). In the analysis using complete cases only, the estimated effect for this interaction term was $\widehat{\eta_7}=0.68$. Therefore, in light of the theory and simulation results and if all assumptions of the analysis are correct, we expect that neglecting this interaction in the sensitivity analysis may somewhat exaggerate the effect of cannabis use on mental health. 

We implemented the three sensitivity analysis approaches as in the simulation study, using the participants’ age $A$ as an auxiliary variable in the implementations of the NAR-SMCFCS and NAR-SMC-stack approaches. Specifically, the ``substantive outcome model’’ considered when implementing these approaches included the auxiliary variable, but this was omitted in the actual substantive outcome used in the analysis, i.e. the outcome model for g-computation (see Section \ref{sec: Discussion}). For the continuous outcome measure considered in the case study, such manipulation is expected to improve the performance of imputation while not resulting in issues due to incompatibility in principle \cite{bartlett2015multiple}. 

The sensitivity analysis approaches were implemented under three possible missingness assumptions: (i) the outcome has a null effect on its own missingness (the non-identifiable part is omitted); (ii) the mean of the outcome is 0.5 units higher in those with missing outcomes than those with observed outcome, conditional on all other variables (the non-identifiable part is $0.5M_Y$); (iii) the conditional mean outcome is 0.3 and 0.5 units higher in those with missing versus observed outcomes in the unexposed and exposed groups, respectively (the non-identifiable part is $0.3M_Y+0.2M_Y X$). We used 10 iterations and generated 100 imputations for each approach, to reduce the Monte Carlo error \cite{white2011multiple}. 

Table \ref{tab: Case.study} presents the results from the case study. As expected, the ACE estimates given by the naïve NARFCS (0.22, 0.26 and 0.29 under the three missingness assumptions, respectively) were all somewhat larger than those given by the proposed approaches under the same missingness assumptions, which were around 0.19, 0.24 and 0.26, respectively. 

Although the differences between approaches in this example were not substantial, the results illustrate the potential importance of compatibility when conducting sensitivity analyses. Suppose that, as is common in practice, the analysis under missingness assumption (i) was conducted using MAR-SMCFCS, while the sensitivity analysis allowing for the outcome to cause its own missingness was implemented using the naïve NARFCS. The results from the case study suggest that this may cause one to over- or under-estimate the impact of the missingness assumptions on ACE estimation. For example, the estimated ACE was 0.19 from the MAR-SMCFCS approach under missingness assumption (i) and 0.26 from the naïve NARFCS approach under missingness assumption (ii). Comparing these two results, one may mistakenly conclude that the ACE estimates are more sensitive to missingness mechanisms than they are in reality: the ACE estimate given by NAR-SMCFCS was 0.24 which is somewhat closer to the 0.19 from the MAR-SMCFCS approach. Although the difference in this case is not consequential, it is easy to see that in more extreme settings neglecting compatibility in the sensitivity analysis could even lead to an opposite conclusion regarding the sensitivity of results, e.g. if the impact of increasing the sensitivity parameter on ACE estimates is in the opposite direction to the bias in ACE estimates due to incompatibility.

\section{Discussion}\label{sec: Discussion}
In this paper, we proposed two approaches for substantive model compatible MI with delta-adjustment for sensitivity analyses to missingness assumptions in the context where an outcome causes its own missingness. Our focus was on the application of the approaches in a causal inference context, but the approaches are applicable more generally.
 
In the causal inference setting, other approaches aside from MI have been studied to tackle scenarios where the ACE is non-recoverable, with emerging work considering acyclic-directed mixed graphs \cite{saadati2019adjustment,bhattacharya2020identification,nabi2020full} and shadow variables \cite{d2010new,wang2014instrumental,ding2014identifiability,miao2016varieties,yang2019causal,sun2021semiparametric}. However, these approaches may be less amenable to practical settings, where it is common to use MI under the assumption that the target estimand is recoverable as a primary analysis, and then conduct sensitivity analysis for assessing departures from this assumption \cite{lee2023assumptions}. Being able to apply an extension of the MI approach used in the primary analysis in the sensitivity analysis facilitates implementation and interpretation of the latter. 

In describing the proposed approaches and in the simulation study, we focused on a specific m-DAG in which the outcome causes its own missingness, a feature that is known to render the ACE non-recoverable. Another type of outcome missingness mechanism that requires sensitivity analysis is when the outcome is associated by its own missingness due to an unmeasured common cause of the outcome and its missingness indicator. This type of missingness mechanism can also be handled using the approaches described here. Other missingness scenarios, such as where an incomplete covariate is associated with its own missingness, may also lead to non-recoverability of the ACE \cite{mohan2014graphical,zhang2024recoverability}. However, standard MI may not result in substantial bias in ACE estimates in these settings and so may not warrant the conduct of sensitivity analyses. For example, a previous study showed that MI (and complete-case analysis) were approximately unbiased in some of the scenarios where the ACE was non-recoverable (when the exposure and confounders caused their own missingness, but the outcome did not influence missingness in any variables) \cite{zhang2024recoverability}. Nonetheless, the proposed approaches could be extended to other settings wherever the delta-adjustment is needed. 

In general, compatibility may be questioned if the substantive model is non-linear or includes interaction or non-linear terms. In this paper, we primarily considered incompatibility in the context of a logistic or a linear regression outcome model with exposure-confounder interactions in the substantive analysis, which is a common occurrence in the causal inference context. The presence of quadratic terms is in the substantive analysis another potential source of incompatibility that has been frequently considered in simulation studies investigating the performance of compatible imputation approaches \cite{bartlett2015multiple,beesley2021stacked}. Given that quadratic terms can be regarded as a type of interaction, i.e. a covariate interacting with itself, the proposed approaches could easily extend to this setting and would be expected to have good performance, although this could be investigated in future research. Cox proportional hazards models, used to model time-to-event outcomes, are another potential source of incompatibility \cite{bartlett2015multiple,beesley2021stacked}, which we did not consider in this study. Incompleteness in a time-to-event outcome is commonly conceptualised as censoring, which is different to full missingness and has different implications for both examining outcome missingness and defining a meaningful causal estimand. It would be valuable for further research to investigate the performance of the proposed approaches in the context of survival analysis. 

One open question for both the SMCFCS and SMC-stack approaches is the treatment of auxiliary variables. In general, two possible approaches are suggested in the MAR-SMCFCS literature for how to incorporate auxiliary variables \cite{bartlett2022auxiliary}. The first strategy is to include the auxiliary variable as a covariate in the substantive outcome model within the imputation procedure and then remove it from the model for substantive analysis. This approach was used in the case study, in which the auxiliary variable was incomplete, and will usually not lead to material incompatibility as the imputation model is larger than the analysis model, unless the link function for the substantive outcome model is not the identity \cite{carpenter2023multiple}. The other strategy is to include auxiliary variables in the imputation models for non-outcome variables but exclude them from the outcome imputation model \cite{bartlett2022auxiliary}. This is the strategy we adopted in the simulation study where the auxiliary was a complete variable. Such a strategy may also be useful in settings where the auxiliary variable should not be included in the outcome imputation model because doing so may cause bias due to the causal structure, e.g. by opening a collider path \cite{thoemmes2014cautious,curnow2023auxiliary}. However, it is unclear whether this strategy violates compatibility when the auxiliary variable has missing data and needs to be imputed \cite{carpenter2023multiple}. In the simulation study, we generated the missingness indicators from the selection model framework rather than the pattern-mixture framework, because this reflected how missingness indicators were caused by the substantive variables as depicted in the m-DAG. As mentioned previously, this means that the outcome imputation model used in all the sensitivity analysis approaches was in some cases incorrectly specified as it was conditional on $M_Y$. Thus, in such settings, the `true' delta values used in the simulations, which were estimated by fitting this model to a large dataset, represent only an approximation to the closest possible pattern-mixture representation.

In this paper, we focussed on delta-adjustment within the pattern-mixture framework. Other sensitivity analysis approaches have been suggested, such as using a selection model framework \cite{heckman1976common}. For example, a sensitivity analysis approach based on the selection model framework was developed for the stacked imputation \cite{beesley2021accounting}. This approach imputes missing values under the MAR assumption and weights the imputations in the stacked dataset by the conditional probability of being missing given the imputed values of incomplete variables \cite{beesley2021accounting}. Future work on developing a compatible implementation for this alternative approach under the selection model framework would be valuable.

In the simulation study, we found that variance estimation in the NAR-SMC-stack approach was downwardly biased. In a supplementary simulation (see Supplementary Material), we observed that this also occurred with the MAR-SMC-stack approach in scenarios where the ACE was equal to the exposure coefficient in a linear outcome regression model and was recoverable. This has also been noted in the importance sampling literature, which shows that when the proposal distribution is close to or proportional to the target distribution, corresponding to missingness patterns II and IV in our settings, the variance estimator may lead to conservative variance estimation in the stacked MI approaches \cite{owen2013importance}. Further research to improve the performance of variance estimators for the stacked approach is needed.

In summary, our findings suggest that both the proposed NAR-SMCFCS and NAR-SMC-stack approaches perform well and that compatibility between the imputation and analysis models is important when conducting sensitivity analysis. Although the target estimand in this paper was the ACE, these approaches should be applicable to other settings. Based on our findings, we recommend that compatible imputation approaches should be implemented when conducting sensitivity analysis with MI in practice.

\section{DECLARATIONS}

\noindent {\bf{Supplementary material}}\\
Supplementary material are available at Statistics in Medicine Journal online.
\\
\noindent {\bf{Ethics approval}}\\
The case study used data from the Victorian Adolescent Health Cohort Study. Data collection protocols were approved by The Royal Children’s Hospital’s Ethics in Human Research Committee. Informed parental consent was obtained for each participant prior to entry.
\\
\noindent {\bf{Funding}}\\
The following authors were supported by National Health and Medical Research Council (NHMRC) Investigator Grants: SGD (ID 2027171), KJL (ID 2017498), MMB (ID 2009572). JWB was supported by UK MRC grant MR/T023953/1. JZ is funded by a University of Melbourne Research Scholarship. The Murdoch Children’s Research Institute is supported by the Victorian Government’s Operational Infrastructure Support Program. The funding bodies did not have any role in the collection, analysis, interpretation or writing of the study.
\\
\noindent {\bf{Data availability}}\\
The data underlying this article will be shared on reasonable request to the corresponding author with the permission of the Chief Investigators of the study, Professor Craig Olsson and Professor Susan Sawyer.\\
\noindent {\bf{Acknowledgements}}\\
The authors would like to thank the Victorian Centre for Biostatistics (ViCBiostat), Causal Inference group, Missing Data group and other members of ViCBiostat for providing feedback in designing and interpreting the simulation study. We also wish to thank the families who participated in the VAHCS, the study research team, and the Principal Investigator of the study, the late Professor George Patton.
\\
\noindent {\bf{Author contributions}}\\
JZ, SGD, JBC, KJL, JWB and MMB conceived the project and designed the study. JZ completed the coding and designed the simulation study, with input from co-authors, and drafted the manuscript. MMB, SGD, JBC, KJL and JWB provided critical input to the manuscript. All of the co-authors read and approved the final version of this paper.
\\
\noindent {\bf{Conflict of interest}}\\
None declared

\begin{sidewaystable}[!ht]
\caption{Descriptive statistics for the analysis variables for the case study, using data from Victorian Adolescent Health Cohort Study ($n=961$)}
\label{tab: Table.one}
\centering
\begin{tabular}{cclccc}
\hline
\multicolumn{3}{c}{}                                        & \multicolumn{2}{c}{Stratified by   exposure\textsuperscript{a}} &              \\
Role       & Label & Variable                               & Unexposed              & Exposed               & Missing (\%) \\ 
\hline
Exposure   & $X$   & Cannabis use in adolescence, Yes                      & 603 (62.7)             & 84 (8.7)              & 28.5            \\
Outcome (continuous)    & $Y$   & Adulthood mental health score \textsuperscript{b}   & -0.11 (1.00)           & 0.48 (0.87)           & 10.4         \\
Outcome (binary) & $Y$ & Adulthood depression \& anxiety, Yes & 101 (17.9)  &  28 (37.8) & 10.4   \\
Confounder & $C_1$  & Parental education (failure to complete high-school), Yes & 206 (34.2)             & 35 (41.7)             & 0            \\
Confounder & $C_2$  & Parental divorce or separation, Yes                  & 94 (15.6)             & 38 (45.2)             & 0            \\
Confounder & $C_3$   & Antisocial behaviour in adolescence, Yes              & 42 (7.0)               & 31 (36.9)             & 0            \\
Confounder & $C_4$  & Adolescent depression \& anxiety, Yes   & 297 (49.3)             & 62 (81.6)             & 12.4         \\
Confounder & $C_5$   & Alcohol use in adolescence, Yes                       & 155 (25.7)             & 65 (87.8)             & 19.3         \\
Auxiliary  & $A$   & Adjusted participant’s age at wave two \textsuperscript{b}       & -0.10 (0.81)           & 0.14 (0.94)           & 8.9          \\ 
Cause of missingness \textsuperscript{c} & $W$ & Adolescent smoking daily, Yes & 100 (16.6)	& 62 (73.8)	& 0 \\
\hline
\end{tabular}

\raggedright
a. For incomplete variables, the descriptive statistics are obtained from the records with available data on the given variable.

b. In standard deviation units, standardised to the overall sample.

c. In simulation studies, variable $W$ is used as an unmeasured common cause of missingness indicators.
\end{sidewaystable}

\clearpage
\begin{sidewaystable}[!ht]
\caption{Simulation study results: coverage probability (\%) for missing data approaches in simple (left panel) and complex (right panel) missingness mechanism scenarios, across a range of outcome scenarios.}
\label{tab: Simu.coverage}
\centering
\begin{tabular}{cc|ccc|ccc}
\hline
\multicolumn{2}{c|}{\multirow{2}{*}{Outcome scenarios}}               & \multicolumn{3}{c|}{simple missingness scenario} & \multicolumn{3}{c}{complex missingness scenario} \\ \cline{3-8} 
\multicolumn{2}{c|}{}                                                 & naïve NARFCS      & NAR-SMCFCS     & NAR-SMC-stack     & naïve NARFCS      & NAR-SMCFCS     & NAR-SMC-stack     \\ \hline
\multicolumn{1}{c|}{\multirow{3}{*}{continuous}} & no interaction     & 93.9\%      & 94.8\%         & 90.9\%            & 95.5\%      & 96.3\%         & 92.8\%            \\
\multicolumn{1}{c|}{}                            & weak interaction   & 94.2\%      & 95.6\%         & 92.1\%            & 95.8\%      & 96.8\%         & 93.1\%            \\
\multicolumn{1}{c|}{}                            & strong interaction & 86.8\%      & 95.5\%         & 92.4\%            & 87.8\%      & 96.0\%         & 93.3\%            \\ \hline
\multicolumn{1}{c|}{\multirow{3}{*}{binary}}     & no interaction     & 94.9\%      & 95.1\%         & 91.6\%            & 95.3\%      & 95.4\%         & 91.5\%                  \\
\multicolumn{1}{c|}{}                            & weak interaction   & 94.9\%      & 95.1\%         & 91.2\%            & 94.2\%      & 94.8\%         & 90.3\%                  \\
\multicolumn{1}{c|}{}                            & strong interaction & 94.9\%      & 94.9\%         & 91.2\%            & 94.3\%      & 94.3\%         & 90.4\%                  \\ \hline
\end{tabular}

\end{sidewaystable}

\begin{table}[!ht]
\caption{Estimates of ACE obtained using missing data methods in the case study ($n=961$)}
\label{tab: Case.study}
\centering
\begin{tabular}{cccc}
\hline
\multicolumn{1}{c|}{Approach}      & Estimate & Standard error & 95\% Confidence interval \\ \hline
\multicolumn{4}{l}{Missingness assumption (i) null}                                                 \\ \hline
\multicolumn{1}{c|}{naïve NARFCS}        & 0.22 & 0.13 & (-0.03, 0.47) \\
\multicolumn{1}{c|}{NAR-SMCFCS}    & 0.19 & 0.14 & (-0.09, 0.47) \\
\multicolumn{1}{c|}{NAR-SMC-stack} & 0.18 & 0.12 & (-0.05, 0.42) \\ \hline
\multicolumn{4}{l}{Missingness assumption (ii) $0.5M_Y$}                                                 \\ \hline
\multicolumn{1}{c|}{naïve NARFCS}        & 0.26 & 0.13 & (0.01, 0.52) \\
\multicolumn{1}{c|}{NAR-SMCFCS}    & 0.24 & 0.13 & (-0.03, 0.50) \\
\multicolumn{1}{c|}{NAR-SMC-stack} & 0.24 & 0.12 & (0.00, 0.47) \\ \hline
\multicolumn{4}{l}{Missingness assumption (iii) $0.3M_Y+0.2M_YX$}                                                 \\ \hline
\multicolumn{1}{c|}{naïve NARFCS}        & 0.29 & 0.12 & (0.04, 0.53) \\
\multicolumn{1}{c|}{NAR-SMCFCS}    & 0.26 & 0.14 & (-0.01, 0.54) \\
\multicolumn{1}{c|}{NAR-SMC-stack} & 0.26 & 0.12 & (0.02, 0.49) \\ \hline
\end{tabular}
\end{table}

\clearpage
\bibliographystyle{unsrtnat}  

\bibliography{article.bib}

\end{document}